\documentclass[aps,pra,twocolumn]{revtex4}
\usepackage{epsfig,amssymb,amsmath,amsthm}
\begin{document}
\title{The Entropy Photon-Number Inequality and its Consequences\footnote{Research supported by the Defense Advanced Research Projects Agency and by the W. M. Keck Foundation Center for Extreme Quantum Information Theory.}} 
\author{Saikat Guha, Baris I. Erkmen, and Jeffrey H. Shapiro}
\affiliation{Massachusetts Institute of Technology, Research Laboratory of Electronics}

\begin{abstract}
Determining the ultimate classical information carrying capacity of electromagnetic waves requires quantum-mechanical analysis to properly account for the bosonic nature of these waves.  Recent work has established capacity theorems for bosonic single-user, broadcast, and wiretap channels, under the presumption of two minimum output entropy conjectures.  Despite considerable accumulated evidence that supports the validity of these conjectures, they have yet to be proven.  Here we show that the preceding minimum output entropy conjectures are simple consequences of an Entropy Photon-Number Inequality, which is a conjectured quantum-mechanical analog of the Entropy Power Inequality from classical information theory.
\end{abstract}

\maketitle

\section{Motivation and History}

The performance of communication systems that rely on electromagnetic wave propagation are ultimately limited by noise of quantum-mechanical origin.  Moreover, high-sensitivity photodetection systems have long been close to or at this noise limit, hence determining the ultimate capacities of lasercom channels, which requires quantum-mechanical analysis, is of immediate relevance.  The most famous channel capacity formula is Shannon's result for the classical additive white Gaussian noise channel.  For a complex-valued channel model in which we transmit $a$ and receive $c = \sqrt{\eta}\,a + \sqrt{1-\eta}\,b$, where $0< \eta < 1$ is the channel's transmissivity and $b$ is a zero-mean, isotropic, complex-valued Gaussian random variable that is independent of $a$, Shannon's capacity is
\begin{equation}
C_{\rm classical} = \ln[1 + \eta\bar{N}/(1-\eta)N]\,\mbox{ nats/use},
\label{Shannon}
\end{equation}
with $E(|a|^2) \le \bar{N}$ and $E(|b|^2) = N$.  
In the quantum version of this channel model, we control the state of an electromagnetic mode with photon annihilation operator $\hat{a}$ at the transmitter, and receive another mode with photon annihilation operator $\hat{c} = \sqrt{\eta}\,\hat{a} + \sqrt{1-\eta}\,\hat{b}$, where $\hat{b}$ is the annihilation operator of a noise mode that is in a zero-mean, isotropic, complex-valued Gaussian state.  For lasercom, if quantum measurements corresponding to ideal optical homodyne or heterodyne detection are employed at the receiver, this quantum channel reduces to a real-valued (homodyne) or complex-valued (heterodyne) additive Gaussian noise channel, from which the following capacity formulas (in nats/use) follow:
\begin{eqnarray}
C_{\rm homodyne} &=& \frac{1}{2}\ln[1+ 4\eta\bar{N}/(2(1-\eta)N + 1)]
\label{homodyne}\\[.12in]
C_{\rm heterodyne} &=& \ln[1+ 2\eta\bar{N}/((1-\eta)N+1)].
\label{heterodyne}
\end{eqnarray}
The + 1 terms in the capacity formulas' noise denominators are quantum noise contributions, so that even when the noise mode $\hat{b}$ is unexcited---in its vacuum state---these capacities remain finite, unlike the situation in 
Eq.~(\ref{Shannon}).  

The classical capacity of the pure-loss bosonic channel---in which the $\hat{b}$ mode is unexcited ($N = 0$)---was shown in \cite{ultcap} to be $C_{\rm pure-loss} = g(\eta\bar{N})$ nats/use, where $g(x) \equiv (x+1)\ln(x+1) - x\ln(x)$ is the Shannon entropy of the Bose-Einstein probability distribution with mean $x$.  This capacity exceeds the $N = 0$ versions of Eqs.~(\ref{homodyne}) and (\ref{heterodyne}), as well as the best known bound on the capacity of photon-number measurement (ideal optical direct detection).  The ultimate capacity of the thermal-noise ($N > 0$) version of this channel was considered in \cite{thermal}, where the lower bound $C_{\rm thermal} \ge g(\eta\bar{N} + (1-\eta)N) - g((1-\eta)N)$ was shown to be the capacity if the thermal channel obeyed a certain minimum output entropy conjecture. This conjecture states that the von Neumann entropy at the output of the thermal channel is minimized when the $\hat{a}$ mode is in its vacuum state.   Several partial results, numerical evidence, and a suite of upper and lower bounds were obtained to support the conjecture \cite{gglms}, but it has yet to be proven. Nevertheless, the preceding lower bound already exceeds Eqs.~(\ref{homodyne}) and (\ref{heterodyne}) as well as the best known bounds on the capacity of direct detection.  

The thermal channel work was followed by a capacity analysis of the bosonic broadcast channel, for which we obtained an inner bound on the capacity region \cite{broadcast}, which we showed to be the capacity region under the presumption of a second minimum output entropy conjecture that was a dual to the first. Both conjectures have been proven if the input states are restricted to be Gaussian, and we have shown that they are equivalent under this input-state restriction.  In unpublished work \cite{wiretap}, we have shown that the second conjecture will establish the privacy capacity of the lossy bosonic channel, as well as its ultimate quantum information carrying capacity. 

The Entropy Power Inequality (EPI) from classical information theory is widely used in coding theorem converse proofs for Gaussian channels.  By analogy with the EPI, we conjecture its quantum version, viz., the Entropy Photon-number Inequality (EPnI).  In this paper we show that the two minimum output entropy conjectures cited above are simple corollaries of the EPnI.  Hence, proving the EPnI would immediately establish key results for the capacities of bosonic communication channels.

\section{Description of the Problem}

\subsection{The Entropy Power Inequality}
Let $\bf X$ and $\bf Y$ be statistically independent, $n$-dimensional, real-valued random vectors that possess differential (Shannon) entropies $h({\bf X})$ and $h({\bf Y})$ respectively. Because a real-valued, zero-mean Gaussian random variable $U$ has differential entropy given by $h(U) = \ln(2\pi e \langle U^2\rangle)$, where the mean-squared value, $\langle U^2\rangle$, is considered to be the \em power\/\rm\ of $U$, the entropy powers of ${\bf X}$ and ${\bf Y}$ are taken to be
\begin{equation}
P({\bf X}) \equiv \frac{e^{h({\bf X})/n}}{2\pi e}\quad\mbox{and}\quad
P({\bf Y}) \equiv \frac{e^{h({\bf Y})/n}}{2\pi e}.
\end{equation}
In this way, an $n$-dimensional, real-valued, random vector $\tilde{\bf X}$ comprised of independent, identically distributed (i.i.d.), real-valued, zero-mean, variance-$P({\bf X})$, Gaussian random variables has differential entropy $h(\tilde{\bf X}) = h({\bf X})$.  We can similarly define an i.i.d. Gaussian random vector $\tilde{\bf Y}$ with differential entropy $h(\tilde{\bf Y}) = h({\bf Y})$. We define a new random vector by 
\begin{equation}
{\bf Z} \equiv \sqrt{\eta}\,{\bf X} + \sqrt{1-\eta}\,{\bf Y}, 
\end{equation}
where $0\le \eta \le 1$.  This random vector has differential entropy $h({\bf Z})$ and entropy power $P({\bf Z})$.  Furthermore, let $\tilde{\bf Z} \equiv \sqrt{\eta}\,\tilde{{\bf X}} + \sqrt{1-\eta}\,\tilde{{\bf Y}}$. Three equivalent forms of the Entropy Power Inequality (EPI), see, e.g., \cite{Rioul}, are then:
\begin{eqnarray}
P({\bf Z}) &\ge& \eta P({\bf X}) + (1-\eta)P({\bf Y})
\label{EPI1}\\[.12in]
h({\bf Z}) &\ge& h(\tilde{\bf Z}) 
\label{EPI2}\\[.12in]
h({\bf Z}) &\ge& \eta h({\bf X}) + (1-\eta) h({\bf Y}).
\label{EPI3}
\end{eqnarray}

\subsection{The Entropy Photon-Number Inequality}
Let $\hat{\boldsymbol a} = [\begin{array}{cccc} \hat{a}_1 & \hat{a}_2 & \cdots & \hat{a}_n\end{array}]$ and $\hat{\boldsymbol b} = [\begin{array}{cccc} \hat{b}_1 & \hat{b}_2 & \cdots & \hat{b}_n\end{array}]$ be vectors of photon annihilation operators for a collection of 2$n$ different electromagnetic field modes of frequency $\omega$ \cite{quantumoptics}. The joint state of the modes associated with $\hat{\boldsymbol a}$ and $\hat{\boldsymbol b}$ is assumed to be the product-state density operator $\hat{\rho}_{\boldsymbol a\boldsymbol b} = \hat{\rho}_{\boldsymbol a}\otimes \hat{\rho}_{\boldsymbol b}$, where $\hat{\rho}_{\boldsymbol a}$ and $\hat{\rho}_{\boldsymbol b}$ are the density operators associated with the $\hat{\boldsymbol a}$ and $\hat{\boldsymbol b}$ modes.  The von Neumann entropies of the $\hat{\boldsymbol a}$ and $\hat{\boldsymbol b}$ modes are $S(\hat{\rho}_{\boldsymbol a}) = -{\rm tr}[\hat{\rho}_{\boldsymbol a}\ln(\hat{\rho}_{\boldsymbol a})]$ and
$S(\hat{\rho}_{\boldsymbol b}) = -{\rm tr}[\hat{\rho}_{\boldsymbol b}\ln(\hat{\rho}_{\boldsymbol b})]$ \cite{footnote2}.  

The thermal state of a mode with annihilation operator $\hat{a}$ has two equivalent forms
\begin{equation}
\hat{\rho}_T = \int\!{\rm d}^2\alpha\, \frac{e^{-|\alpha|^2/N}}{\pi N}\,|\alpha\rangle \langle \alpha|,
\label{cohstateform}
\end{equation}
and 
\begin{equation}
\hat{\rho}_T = \sum_{i=0}^\infty \frac{N^i}{(N+1)^{i+1}}\,|i\rangle \langle i|,
\label{numberstateform}
\end{equation}
where $N = \langle \hat{a}^\dagger \hat{a}\rangle$ is the average photon number.  In Eq.~(\ref{cohstateform}), $|\alpha\rangle$ is the coherent state of amplitude $\alpha$, i.e., it satisfies $\hat{a}|\alpha\rangle = \alpha|\alpha\rangle$, for $\alpha$ a complex number. In Eq.~(\ref{numberstateform}), $|i\rangle$ is the $i$-photon state, i.e., it satisfies
$\hat{N}|i\rangle = i|i\rangle$, for $i = 0, 1, 2, \ldots$, with $\hat{N} \equiv \hat{a}^\dagger\hat{a}$ being the photon number operator.  Physically, Eq.~(\ref{cohstateform}) says that the thermal state is an isotropic Gaussian mixture of coherent states.  Equation~(\ref{numberstateform}), on the other hand, says that the thermal state is a Bose-Einstein mixture of number states.  From Eq.~(\ref{numberstateform}) we immediately have that $S(\hat{\rho}_T) = g(N)$, because the photon-number states are orthonormal \cite{footnote3}.
Note that $g(N)$, for $N \ge 0$, is a non-negative, monotonically increasing, concave function of $N$.  Thus it has an inverse, $g^{-1}(S)$, for $S \ge 0$, that is non-negative, monotonically increasing, and convex.

The entropy photon-numbers of the density operators $\hat{\rho}_{\boldsymbol a}$ and $\hat{\rho}_{\boldsymbol b}$ are defined as follows:
\begin{equation}
N(\hat{\rho}_{\boldsymbol a})\equiv g^{-1}(S(\hat{\rho}_{\boldsymbol a})/n)\, \mbox{ and }\,
N(\hat{\rho}_{\boldsymbol b})\equiv g^{-1}(S(\hat{\rho}_{\boldsymbol b})/n).
\end{equation}
Thus, if
$\hat{\rho}_{\tilde{\boldsymbol a}} \equiv \bigotimes_{i=1}^n \hat{\rho}_{T_{a_i}}$ and
$\hat{\rho}_{\tilde{\boldsymbol b}} \equiv \bigotimes_{i=1}^n \hat{\rho}_{T_{b_i}}$,
where $\hat{\rho}_{T_{a_i}}$ is the thermal state of average photon  number $N(\hat{\rho}_{\boldsymbol a})$ for the $\hat{a}_i$ mode and $\hat{\rho}_{T_{b_i}}$ is the thermal state of average photon number $N(\hat{\rho}_{\boldsymbol b})$ for the $\hat{b}_i$ mode, then we have $S(\hat{\rho}_{\tilde{\boldsymbol a}}) = S(\hat{\rho}_{\boldsymbol a})$ and $S(\hat{\rho}_{\tilde{\boldsymbol b}}) = S(\hat{\rho}_{\boldsymbol b})$. We define a new vector of photon annihilation operators,  
$\hat{\boldsymbol c} = [\begin{array}{cccc} \hat{c}_1 & \hat{c}_2 & \cdots & \hat{c}_n\end{array}]$,
by
\begin{equation}
\hat{\boldsymbol c} \equiv \sqrt{\eta}\,\hat{\boldsymbol a} + \sqrt{1-\eta}\,\hat{\boldsymbol b},
\quad\mbox{for $0\le \eta \le 1$,}
\label{beamsplitter_def}
\end{equation}
and use $\hat{\rho}_{\boldsymbol c}$ to denote its density operator.
This is equivalent to saying that $\hat{c}_i$ is the output of a lossless beam splitter whose inputs, $\hat{a}_i$ and $\hat{b}_i$, couple to that output with transmissivity $\eta$ and reflectivity $1-\eta$, respectively.  

We can now state two equivalent forms of our conjectured Entropy Photon-Number Inequality (EPnI) \cite{footnote4}:
\begin{eqnarray}
N(\hat{\rho}_{\boldsymbol c})&\ge& \eta N(\hat{\rho}_{\boldsymbol a})+ 
(1-\eta)N(\hat{\rho}_{\boldsymbol b})\label{EPnI1}\\[.12in]
S(\hat{\rho}_{\boldsymbol c}) &\ge & S(\hat{\rho}_{\tilde{\boldsymbol c}}),
\label{EPnI2}
\end{eqnarray}
where $\hat{\rho}_{\tilde{\boldsymbol c}} \equiv \bigotimes_{i=1}^n \hat{\rho}_{T_{c_i}}$ with $\hat{\rho}_{T_{c_i}}$ being the thermal state of mean photon  number ${\eta}N(\hat{\rho}_{\boldsymbol a})+(1-{\eta})N(\hat{\rho}_{\boldsymbol b})$ for $\hat{c}_i$. 

\section{Prior Work and Discussion}
By analogy with the classical EPI, we might expect there to be a third equivalent form of the quantum EPnI, viz.,
\begin{equation}
S(\hat{\rho}_{\boldsymbol c}) \ge \eta S(\hat{\rho}_{\boldsymbol a}) + (1-\eta) S(\hat{\rho}_{\boldsymbol b}).
\label{EPnI3}
\end{equation}
It is easily shown that Eq.~(\ref{EPnI1}) implies Eq.~(\ref{EPnI3}) \cite{footnote5}, but we have not been able to prove the converse.  Indeed, we suspect that the converse might be false. More important than whether or not (\ref{EPnI3}) is equivalent to Eq.~(\ref{EPnI1}) and Eq.~(\ref{EPnI2}), is the role of the EPnI in proving classical information capacity results for bosonic channels.  In particular, the EPnI provides simple proofs of the following two minimum output entropy conjectures.  These conjectures are important because proving minimum output entropy conjecture~1 also proves the conjectured capacity of the thermal-noise channel \cite{thermal}, and proving minimum output entropy conjecture~2 also proves the conjectured capacity region of the bosonic broadcast channel \cite{broadcast}. 

{\bf{Minimum Output Entropy Conjecture 1 ---}}
Let ${\boldsymbol a}$ and ${\boldsymbol b}$ be $n$-dimensional vectors of annihilation operators, with joint density operator $\hat{\rho}_{\boldsymbol a\boldsymbol b} = (|\psi\rangle_{\boldsymbol a}{}_{\boldsymbol a}\langle \psi|)\otimes \hat{\rho}_{\boldsymbol b}$, where $|\psi\rangle_{\boldsymbol a}$ is an arbitrary zero-mean-field pure state of the ${\boldsymbol a}$ modes and  $\hat{\rho}_{\boldsymbol b} = \bigotimes_{i=1}^n\hat{\rho}_{T_{b_i}}$ with $\hat{\rho}_{T_{b_i}}$ being the $\hat{b}_i$ mode's thermal state of average photon number $K$.  
Define a new vector of photon annihilation operators,  
$\hat{\boldsymbol c} = [\begin{array}{cccc} \hat{c}_1 & \hat{c}_2 & \cdots & \hat{c}_n\end{array}]$,
by~\eqref{beamsplitter_def}
and use $\hat{\rho}_{\boldsymbol c}$ to denote its density operator and $S(\hat{\rho}_{\boldsymbol c})$ to denote its von Neumann entropy.  Then choosing $|\psi\rangle_{\boldsymbol a}$ to be the $n$-mode vacuum state minimizes $S(\hat{\rho}_{\boldsymbol c})$.  

{\bf{Minimum Output Entropy Conjecture 2 ---}}
Let ${\boldsymbol a}$ and ${\boldsymbol b}$ be $n$-dimensional vectors of annihilation operators, as in the previous section, with joint density operator $\hat{\rho}_{\boldsymbol a\boldsymbol b} = (|\psi\rangle_{\boldsymbol a}{}_{\boldsymbol a}\langle \psi|)\otimes \hat{\rho}_{\boldsymbol b}$, where $|\psi\rangle_{\boldsymbol a} = \bigotimes_{i=1}^n|0\rangle_{a_i}$ is the $n$-mode vacuum state and 
$\hat{\rho}_{\boldsymbol b}$ has von Neumann entropy $S(\hat{\rho}_{\boldsymbol b}) = ng(K)$ for some $K \ge 0$.    
Define a new vector of photon annihilation operators,  
$\hat{\boldsymbol c} = [\begin{array}{cccc} \hat{c}_1 & \hat{c}_2 & \cdots & \hat{c}_n\end{array}]$,
by~\eqref{beamsplitter_def}
and use $\hat{\rho}_{\boldsymbol c}$ to denote its density operator and $S(\hat{\rho}_{\boldsymbol c})$ to denote its von Neumann entropy.  Then choosing $\hat{\rho}_{\boldsymbol b} = \bigotimes_{i=1}^n\hat{\rho}_{T_{b_i}}$ with $\hat{\rho}_{T_{b_i}}$ being the $\hat{b}_i$ mode's thermal state of average photon number $K$ minimizes  $S(\hat{\rho}_{\boldsymbol c})$.

To see that the EPnI encompasses both of the preceding minimum output entropy conjectures is our final task in this paper.  We begin by using the premise of conjecture~1 in Eq.~(\ref{EPnI1}).  Because the $\hat{\boldsymbol a}$ modes are in a pure state, we get $S(\hat{\rho}_{\boldsymbol a})= 0$ and hence
the EPnI tells us that
\begin{equation}
N(\hat{\rho}_{\boldsymbol c}) \ge  (1-\eta)N(\hat{\rho}_{\boldsymbol b}) = (1-\eta)K.
\end{equation}
Taking $g(\cdot)$ on both sides of this inequality, we get $S(\hat{\rho}_{\boldsymbol c})/n \ge g[(1-\eta)K]$. 
But, if $|\psi\rangle_{\boldsymbol a}$ is the $n$-mode vacuum state, we can easily show that 
$\hat{\rho}_{\boldsymbol c} = \bigotimes_{i=1}^n\hat{\rho}_{T_{c_i}}$,
with $\hat{\rho}_{T_{c_i}}$ being the $\hat{c}_i$ mode's thermal state of average photon number $(1-\eta)K$.  Thus, when $|\psi\rangle_{\boldsymbol a}$ is the $n$-mode vacuum state we get
$S(\hat{\rho}_{\boldsymbol c}) = ng[(1-\eta)K]$, which completes the proof.   

Next, we apply the premise of conjecture~2 in Eq.~(\ref{EPnI1}).  Once again, the $\hat{\boldsymbol a}$ modes are in a pure state, so we get
\begin{equation}
N(\hat{\rho}_{\boldsymbol c}) \ge  (1-\eta)N(\hat{\rho}_{\boldsymbol b}) = (1-\eta)K,
\end{equation}
and hence $S(\hat{\rho}_{\boldsymbol c})/n \ge g[(1-\eta)K]$. But, taking $\hat{\rho}_{\boldsymbol b} = \bigotimes_{i=1}^n\hat{\rho}_{T_{b_i}}$, with $\hat{\rho}_{T_{b_i}}$ being the $\hat{b}_i$ mode's thermal state of average photon number $K$, satisfies the premise of minimum output entropy conjecture~2 and implies that $\hat{\rho}_{\boldsymbol c} = \bigotimes_{i=1}^n\hat{\rho}_{T_{c_i}}$, with $\hat{\rho}_{T_{c_i}}$ being the $\hat{c}_i$ mode's thermal state of average photon number $(1-\eta)K$.  In this case we have $S(\hat{\rho}_{\boldsymbol c}) = ng[(1-\eta)K]$, which completes the proof.


\begin{thebibliography}{2}
\bibitem{ultcap} V. Giovannetti, S. Guha, S. Lloyd, L. Maccone, J. H. Shapiro, and H. P. Yuen, Phys. Rev. Lett. {\bf 92}, 027902 (2004).
\bibitem{thermal} V. Giovannetti, S. Guha, S. Lloyd, L. Maccone, J. H. Shapiro,
B. J. Yen, and H. P. Yuen, in \em Quantum Information, Statistics,
Probability\/\rm, edited by O. Hirota  (Rinton Press, Princeton, NJ,
2004), pp. 90--101.
\bibitem{gglms} V. Giovannetti, S. Guha, S. Lloyd, L. Maccone, and J. H. Shapiro, Phys. Rev. A {\bf 70}, 032315 (2004).
\bibitem{broadcast} S. Guha, J. H. Shapiro, and B. I. Erkmen, Phys. Rev. A {\bf 76}, 032303 (2007).
\bibitem{wiretap} S. Guha and J. H. Shapiro, {\em{Unpublished notes}}, (2007).

\bibitem{Rioul}O. Rioul, ``Information theoretic proofs of entropy power inequalities,'' arxiv cs.IT/0704.175
\bibitem{quantumoptics} L. Mandel and E. Wolf, \em Optical Coherence and Quantum Optics\/\rm, (Cambridge University Press, Cambridge, 1995).
\bibitem{footnote2} A density operator is Hermitian, with eigenvalues that form a probability distribution.  Thus, the von Neumann entropy of a density operator $\hat{\rho}$ is the Shannon entropy of its eigenvalues.
\bibitem{footnote3}  The coherent states, $\{|\alpha\rangle\}$, are \em not\/\rm\ orthonormal, but rather overcomplete. 
\bibitem{footnote4} To show that (\ref{EPnI1}) implies (\ref{EPnI2}), assume (\ref{EPnI1}) is true:
\begin{eqnarray}
N(\hat{\rho}_{\boldsymbol c})&\ge& \eta N(\hat{\rho}_{\boldsymbol a})+ 
(1-\eta)N(\hat{\rho}_{\boldsymbol b})\\[.12in]
&=& \eta N(\hat{\rho}_{\tilde{\boldsymbol a}}) + 
(1-\eta) N(\hat{\rho}_{\tilde{\boldsymbol b}})
\label{deriv12}
\end{eqnarray}
Now, if $\hat{\rho}_{\tilde{\boldsymbol a}\tilde{\boldsymbol b}} = \hat{\rho}_{\tilde{\boldsymbol a}} \otimes 
\hat{\rho}_{\tilde{\boldsymbol b}}$ is the joint density operator of the $\hat{\boldsymbol a}$ and $\hat{\boldsymbol b}$ modes, we find that the state of the $\hat{\boldsymbol c}$ modes is 
$\hat{\rho}_{\tilde{\boldsymbol c}} \equiv \bigotimes_{i=1}^n 
\hat{\rho}_{T_{c_i}}$,
where $\hat{\rho}_{T_{c_i}}$ is a thermal state with average photon number given by
$N(\hat{\rho}_{\tilde{\boldsymbol c}}) = \eta N(\hat{\rho}_{\tilde{\boldsymbol a}}) + (1-\eta) N(\hat{\rho}_{\tilde{\boldsymbol b}})$,
so that $S(\hat{\rho}_{\tilde{\boldsymbol c}}) = ng[N(\hat{\rho}_{\tilde{\boldsymbol c}})]$.  
Thus, from (\ref{deriv12}) we get $N(\hat{\rho}_{\boldsymbol c})\ge 
N(\hat{\rho}_{\tilde{\boldsymbol c}}) =  g^{-1}(S(\hat{\rho}_{\tilde{\boldsymbol c}})/n)$. Taking $g(\cdot)$ of both sides of this inequality completes the proof.

To show that (\ref{EPnI2}) implies (\ref{EPnI1}), assume (\ref{EPnI2}) is true:
\begin{eqnarray}
N(\hat{\rho}_{\boldsymbol c})&=& g^{-1}(S(\hat{\rho}_{\boldsymbol c})/n) \nonumber \\[.12in] 
&\ge & g^{-1}(S(\hat{\rho}_{\tilde{\boldsymbol c}})/n)  = 
g^{-1}[g(\eta N(\hat{\rho}_{\tilde{\boldsymbol a}}) + (1-\eta) N(\hat{\rho}_{\tilde{\boldsymbol b}}))] \nonumber \\[.12in]
&=& \eta N(\hat{\rho}_{\tilde{\boldsymbol a}}) + (1-\eta) N(\hat{\rho}_{\tilde{\boldsymbol b}}) \nonumber \\[.12in]
&=& \eta N(\hat{\rho}_{\boldsymbol a})+ (1-\eta) N(\hat{\rho}_{\boldsymbol b}),
\end{eqnarray}
where the inequality is due to $g^{-1}(S)$ being a monotonically increasing function of $S$, and the proof is complete.

\bibitem{footnote5}Assume that  (\ref{EPnI1}) is true. We then have that $N(\hat{\rho}_{\boldsymbol c}) \ge \eta N(\hat{\rho}_{\boldsymbol a})+ (1-\eta)N(\hat{\rho}_{\boldsymbol b})$, so that
\begin{eqnarray}
S(\hat{\rho}_{\boldsymbol c}) &=& ng[N(\hat{\rho}_{\boldsymbol c})] \ge ng[\eta N(\hat{\rho}_{\boldsymbol a}) + (1-\eta) N(\hat{\rho}_{\boldsymbol b})] \\[.12in] 
&\ge& \eta ng[N(\hat{\rho}_{\boldsymbol a})] + (1-\eta)ng[N(\hat{\rho}_{\boldsymbol b})] \\[.12in]
&=& \eta S(\hat{\rho}_{\boldsymbol a}) + (1-\eta)S(\hat{\rho}_{\boldsymbol b}),
\end{eqnarray}
where the second inequality follows from $g(N)$ being concave, and the proof is complete.
\end{thebibliography}
\end{document}